\documentstyle[12pt]{article}
\title{Short Proof of Jacobi's Identity for Poisson Brackets}
\author{ \sc{ Nivaldo A. Lemos}\\
\small\sf{Instituto de F\'{\i}sica - Universidade Federal Fluminense} \\
\small\sf{ Av. Litor\^anea  s/n, 24210-340 Boa Viagem, Niter\'oi, RJ, Brazil}\\
\small\sf{e-mail: nivaldo@if.uff.br}}  
\date{}
\begin{document}
\pagestyle{myheadings} 
\baselineskip 18pt
%\markboth{L.V.Belvedere  }{Supersymmetric Liouville Theory}
\maketitle

\vspace{2cm}

\begin{abstract}  

Making use of the theory of infinitesimal canonical transformations, a concise  proof is given of Jacobi's identity for Poisson brackets.
\end{abstract}
\newpage

According to Goldstein$^1$ ``there seems to be no simple way of proving Jacobi's identity for the Poisson bracket without lengthy algebra." This impression appears to be shared by other authors, who either also explicitly do the lengthy algebra$^{2-5}$ or leave the tedious work to the reader.$^{6,7}$ The purpose of this note is to show that, contrary to this widespread belief, there is an extremely short proof of Jacobi's identity, based on the theory of infinitesimal canonical transformations, with virtually no algebra.

Let $\, A(q,p)\,$ and $\, B(q,p)\,$ be any two dynamical variables and consider an infinitesimal canonical transformation generated by 
$\, C(q,p)$. On the one hand, Goldstein's Eq.(9-103) with $\, u= \{A,B\}\,$ yields

$$\delta \{A,B\} = \epsilon\, \{ \{A,B\} ,C \}
\,\,\, . \eqno(1)$$ 
\\ 
On the other hand,  since the Poisson bracket $\,\{A,B\} \,$ does not depend on the  canonical variables chosen for its computation, its change is due only to the  variations of  $\, A\,$ and 
$\, B$, so that 

$$\delta \{A,B\} =  \{ \delta A,B\} +  \{A,\delta B\}
\,\,\, , \eqno(2)$$ 
\\ 
whence

$$\delta \{A,B\} = \epsilon \{\{A,C\}, B\} + \epsilon \{A,\{ B, C\}\}
\,\,\, . \eqno(3)$$ 
\\
Comparing (1) with (3) and making  simple rearrangements one finds

$$ \{ \{A,B\} ,C \} + \{\{C,A\},B\} +  \{\{B, C\} , A\} = 0
\,\,\, , \eqno(4)$$ 
\\
which is Jacobi's identity.

\newpage
\centerline{\large\bf References}
\vspace{.5cm}
\begin{description}
\item{1.} H. Goldstein, {\it Classical Mechanics} (Addison-Wesley, Reading, MA, 1980), 2nd ed., p. 399.

\item{2.} L. A. Pars, {\it A Treatise on Analytical Dynamics} (Ox Bow, Woodbridge, CT, 1965), p. 429.

\item{3.} F. Gantmacher, {\it Lectures in Analytical Mechanics} (Mir, Moscow, 1970), pp. 84-85.

\item{4.} E. A. Desloge, {\it Classical Mechanics} (Robert E. Krieger, Malabar, FL, 1982), pp. 824-825.

\item{5.} F. Scheck, {\it  Mechanics $-$ From Newton's Laws to Deterministic Chaos} (Springer, Berlin, 1994), p. 132.

\item{6.} D. ter Haar, {\it Elements of Hamiltonian Mechanics} (Pergamon, Oxford, 1963), p. 107.

\item{7.} N. M. J. Woodhouse, {\it Introduction to Analytical Dynamics} (Clarendon, Oxford, 1987), p. 121.

 \end{description}
 \end{document}